\documentclass[letterpaper]{jpconf}
\usepackage{graphicx}
\usepackage{slashed}
\begin{document}
\title{Dynamical Mass Generation and Confinement in Maxwell-Chern-Simons Planar Quantum Electrodynamics}

\author{S S\'anchez Madrigal$^{1}$, C P Hofmann$^{2}$ and A Raya$^{1}$, }

\address{$^1$Instituto de F\'{\i}sica y Matem\'aticas,
Universidad Michoacana de San Nicol\'as de Hidalgo, Edificio C-3, Ciudad Universitaria, Morelia, Michoac\'an
58040, M\'exico.}
\address{$^2$Facultad de Ciencias, Universidad de Colima, Bernal D\'{\i}az del Castillo 340, Colima, Colima
28045, M\'exico.}

\ead{saul@ifm.umich.mx, christoph@ucol.mx, raya@ifm.umich.mx,}

\begin{abstract}
We study the non-perturbative phenomena of Dynamical Mass Generation and Confinement by truncating at the
non-perturbative level the Schwinger-Dyson equations in Maxwell-Chern-Simons planar quantum electrodynamics.
We obtain numerical solutions for the fermion propagator in Landau gauge within the so-called rainbow
approximation. A comparison with the ordinary theory without the Chern-Simons term is presented.
\end{abstract}

\section{Motivation}

Dynamical Mass Generation (DMG) and Confinement are two emergent phenomena of Quantum Chromodynamics (QCD)
responsible for the nature of the hadronic spectrum. These can be studied by means of the corresponding
Schwinger-Dyson Equations (SDE), as an alternative to the formulation of QCD on the lattice. The SDEs form
an infinite tower of relations among the Green's functions of a given quantum field theory, whose nature is
genuinely non-perturbative. We choose to study the SDE in a simplified model that shares with QCD the
phenomena of Confinement and DMG, namely, quantum electrodynamics in the plane, or QED$_3$. Applications of
QED$_3$ also extend to condensed matter systems, including high-T$_c$ superconductivity, the quantum Hall
effect and, very recently, graphene. Moreover, it is interesting to include the Chern-Simons (CS) term for
the gauge field, which explicitly breaks parity and induces a topological mass for the photons. 

Incorporating the CS term in studies of DMG in QED$_3$ have found that when the CS coefficient reaches a
critical value, chiral symmetry is restored in a first order phase transition~\cite{KondMarPRL,KondMarPRD}.
In this work, we are interested in the effect of the CS term regarding confinement in quenched QED$_3$. 
In Sec. 2 we present our model with a reducible representation of (2+1)-dimensional quantum electrodynamics
with a CS term. Solutions to the gap equation in the rainbow approximation are presented in Sec. 3. The
question of confinement in QED$_3$ with a CS term is addressed in Sec.4. Finally, in Sec. 5, we present our
conclusions.

\section{Model}

We study quantum electrodynamics in (2+1)-dimensions. For fermions, there exist two irreducible
representations of the Euclidean Clifford algebra $\{ \gamma_\mu, \gamma_\nu\}=2\delta_{\mu\nu}$. In any of
these representations, it is impossible to define chiral symmetry~\cite{Appel,Pisarski,Semenoff}.
Furthermore, the fermion mass term, regardless its origin, is parity ${\cal P}$ non invariant. One still can
construct a parity-preserving Lagrangian considering two different species with a relative sign between
their masses~\cite{Shimizu,Chuy}. As a result, two chiral transformations can be defined~\cite{Chuy}. These
two species are conveniently merged into a single four-component spinor, making use of a reducible
representation of the Dirac matrices.

As compared with its four-dimensional counterpart, only three Dirac matrices are required to describe the
dynamics of planar fermions. Once we have selected a set of matrices to write down the Dirac equation, say
$\{\gamma_0,\gamma_1,\gamma_2\}$, two anti-commuting gamma matrices, namely, $\gamma_3$ and $\gamma_5$
remain unused. Hence the corresponding massless Dirac Lagrangian is invariant under the chiral-like
transformations $\psi\to e^{i\alpha\gamma_3}\psi$ and $\psi\to e^{i\beta\gamma_5}\psi$, that is, it is invariant
under a global $U(2)$ symmetry with generators $1$, $\gamma_3$, $\gamma_5$ and $[\gamma_3, \gamma_5]$,
corresponding to the interchange of fermion species. This symmetry is broken by an ordinary mass term
$m_e\bar\psi\psi$. The order parameter of the symmetry breaking is the condensate
$\langle \bar\psi\psi \rangle_e=\langle 0|\bar\psi\psi |0\rangle$. There exists a second mass term which
is invariant under the "chiral" transformations, referred to as the Haldane mass term
$ m_o \bar\psi \tau \psi$ ~\cite{Haldane} with $\tau=[\gamma_3,\gamma_5]/2=diag(I,-I)$, which is associated
with the condensate $\langle \bar\psi\psi \rangle_o=\langle 0|\bar\psi\tau\psi |0\rangle$. The ordinary mass
term is even under parity $\cal P$ transformations, but the Haldane mass term is not. We use the
subscripts $e$ for even and $o$ for odd in quantities of the model.
The Haldane mass term radiatively induces a parity-odd 
contribution into the vacuum polarization,
which can be traced back to an induced Chern-Simons interaction
\begin{equation}
{\cal L}_{CS}=-\frac{i\theta}{4}\varepsilon_{\mu\nu\rho}A_\mu F_{\nu\rho}\;.\label{csl}
\end{equation}
Such a term is parity non invariant, and despite the fact that it is not manifestly gauge invariant, the
corresponding action respects gauge symmetry. The parameter $\theta$ induces a topological mass for the
photons. Thus, the model we consider is described by the Lagrangian 
\begin{eqnarray}
{\cal L}&=& \bar\psi (i\slashed{\partial}+e\slashed{A}+m_e+\tau m_o)\psi
+\frac{1}{4}F_{\mu\nu}F_{\mu\nu}
+\frac{1}{2\xi}(\partial_\mu A_\mu)^2-\frac{i\theta}{4}\varepsilon_{\mu\nu\rho}A_\mu F_{\nu\rho}.\label{mcslag}
\end{eqnarray}
There are many condensed matter systems which can be described by this Lagrangian, for which the physical
origin of the masses depends on the underlying system, including high-T$_c$ superconductivity, the quantum
Hall effect and, very recently, graphene.

Written in this form, neither $m_e$ nor $m_o$ correspond to poles of the fermion propagator. We introduce
the chiral projectors $\chi_\pm = (1\pm \tau)/2 $ which have the properties
$\chi_\pm^2=\chi_\pm\;, \chi_+\chi_-=0\;, \chi_++\chi_-=1$. These allow us to define the right-handed
$\psi_+$ and left-handed $\psi_-$ fermion fields as $\psi_\pm=\chi_\pm \psi$. The $\chi_\pm$ project the
upper and lower two component spinors (fermion species) out of the four-component $\psi$, such that the
fermion sector of the Lagrangian~(\ref{mcslag}) can be cast in the form
\begin{equation}
{\cal L}_F= \bar\psi_+ (i\slashed{\partial}-m_+)\psi_++\bar\psi_- (i\slashed{\partial}-m_-)\psi_-,
\label{fermchi}
\end{equation}
with $m_\pm=m_e\pm m_o$. This Lagrangian describes two fermion species of physical masses $m_+$ and $m_-$,
respectively. These masses break chiral symmetry and parity at the same time. Moreover, the effect of the
parity-violating mass is seen to remove the mass degeneracy between the two species. In what follows, we
will be interested in the analytical properties of the dynamically generated fermion propagator associated
with these physical masses. The analytical structure of the propagator can be studied from the corresponding
Schwinger-Dyson equation, which corresponds to the expression
\begin{eqnarray}
S_F(p)^{-1}&=&S_F^{(0)}(p)^{-1}
+\ e^2 \int \frac{d^3k}{(2\pi)^3} \Gamma^\mu(k, p) S_F(k) \gamma^\nu \Delta_{\mu\nu}(k-p)\;,\label{SDE}
\end{eqnarray}
where $\Gamma^\mu(k,p)$ and $\Delta_{\mu\nu}(k-p)$ are, respectively, the full fermion-photon vertex and the
full photon propagator, which obey themselves their own SDE, and $S_F(p)$ and $S_F^{(0)}(p)$ stand for the
full and bare fermion propagator. In QED$_3$, the coupling $e^2$ has mass-dimension one. Moreover, since the
theory is super-renormalizable, $e^2$ becomes the natural scale of massless QED$_3$, which is directly
connected to the scales of confinement and dynamical chiral symmetry breaking. In this work, we write the
relevant mass scales in units of $e^2=1$.
 
From the Lagrangian~(\ref{mcslag}), we observe that inverse fermion propagator has the form
\begin{equation}
S_F^{-1}(p)= A_e(p)\slashed{p} +A_o(p) \tau \slashed{p}-B_e(p)-B_o(p)\tau\;.
\end{equation}
The scalar functions $A_{e,o}(p)$ and $B_{e,o}(p)$ can be expressed in terms of the fermion wavefunction
renormalizations $F_{e,o}(p)$ and the mass function $M_{e,o}(p)$ as $A_{e,o}(p)=1/F_{e,o}(p)$ and
$B_{e,o}(p)=M_{e,o}(p)/F_{e,o}(p)$ both in the even and odd sectors. The bare propagator corresponds to the
values $A_e^{(0)}(p)=1,\,A_o^{(0)}(p)=0,\,B_e^{(0)}(p)=m_e,\,B_o^{(0)}(p)=m_o$. Rather than working with parity
eigenstates, we find it convenient to work with the chiral Lagrangian~(\ref{fermchi}). The chiral
decomposition of the fermion propagator becomes
\begin{eqnarray}
S_F(p)&=& -\frac{A_+ (p)\slashed{p}+B_+(p)}{A_+^2(p)p^2+B_+^2(p)}\chi_+ 
- \frac{A_- (p)\slashed{p}+B_-(p)}{A_-^2(p)p^2+B_-^2(p)}\chi_- 
\nonumber\\
&\equiv& -[\sigma_+^V(p) \slashed{p}+\sigma_+^S(p)] \chi_+
-[\sigma_-^V(p) \slashed{p}+\sigma_-^S(p)] \chi_-
\;,\label{fpp}
\end{eqnarray}
where $K_\pm=K_e\pm K_o$ for all relevant even and odd quantities of the model. The inverse
transformations are simply $K_e=(K_++K_-)/2$ and $K_o=(K_+-K_-)/2$. In this basis, the bare quantities are
$A_\pm^{(0)}(p)=1$ and $B_\pm^{(0)}(p)=m_\pm$. On the other hand, the photon propagator associated with the
Lagrangian (\ref{mcslag}) in the Landau gauge, $\xi=0$, takes the form
\begin{equation}
\Delta_{\mu\nu}^{(0)}(q)=\frac{1}{q^2+\theta^2}\left(\delta_{\mu\nu}-
\frac{q_\mu q_\nu}{q^2} \right)-\frac{\varepsilon_{\mu\nu\rho}q_\rho \theta}{q^2(q^2+\theta^2)}\;.
\end{equation}

\section{Gap Equation}

We consider the rainbow-ladder truncation scheme, in which one makes the replacements
$\Gamma^\mu \to \gamma^\mu$ and $\Delta_{\mu\nu} \to \Delta_{\mu\nu}^{(0)}$.
\begin{figure}
\begin{center}
\includegraphics[width=0.35\textwidth, angle=270]{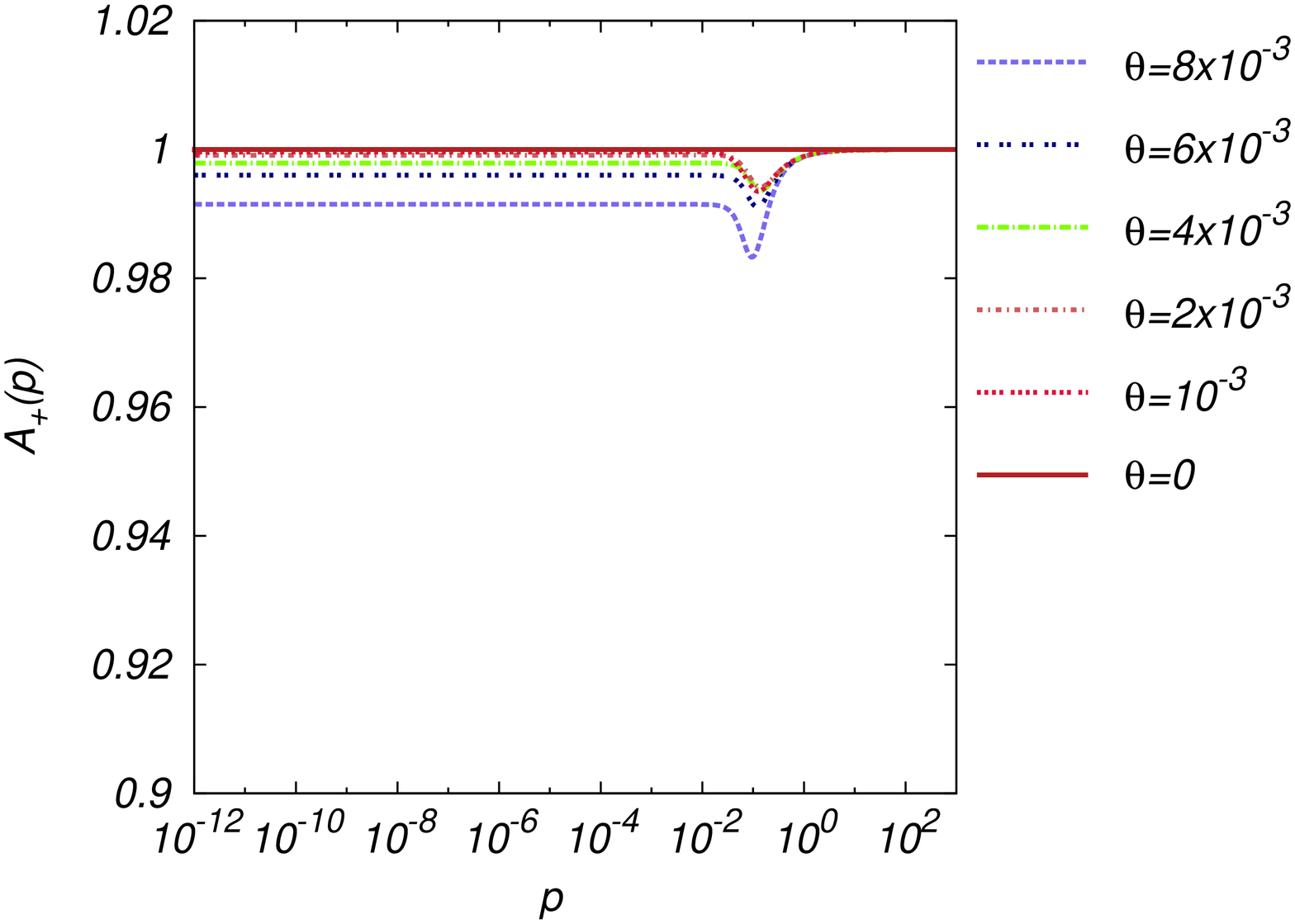}\includegraphics[width=0.35\textwidth,
angle=270]{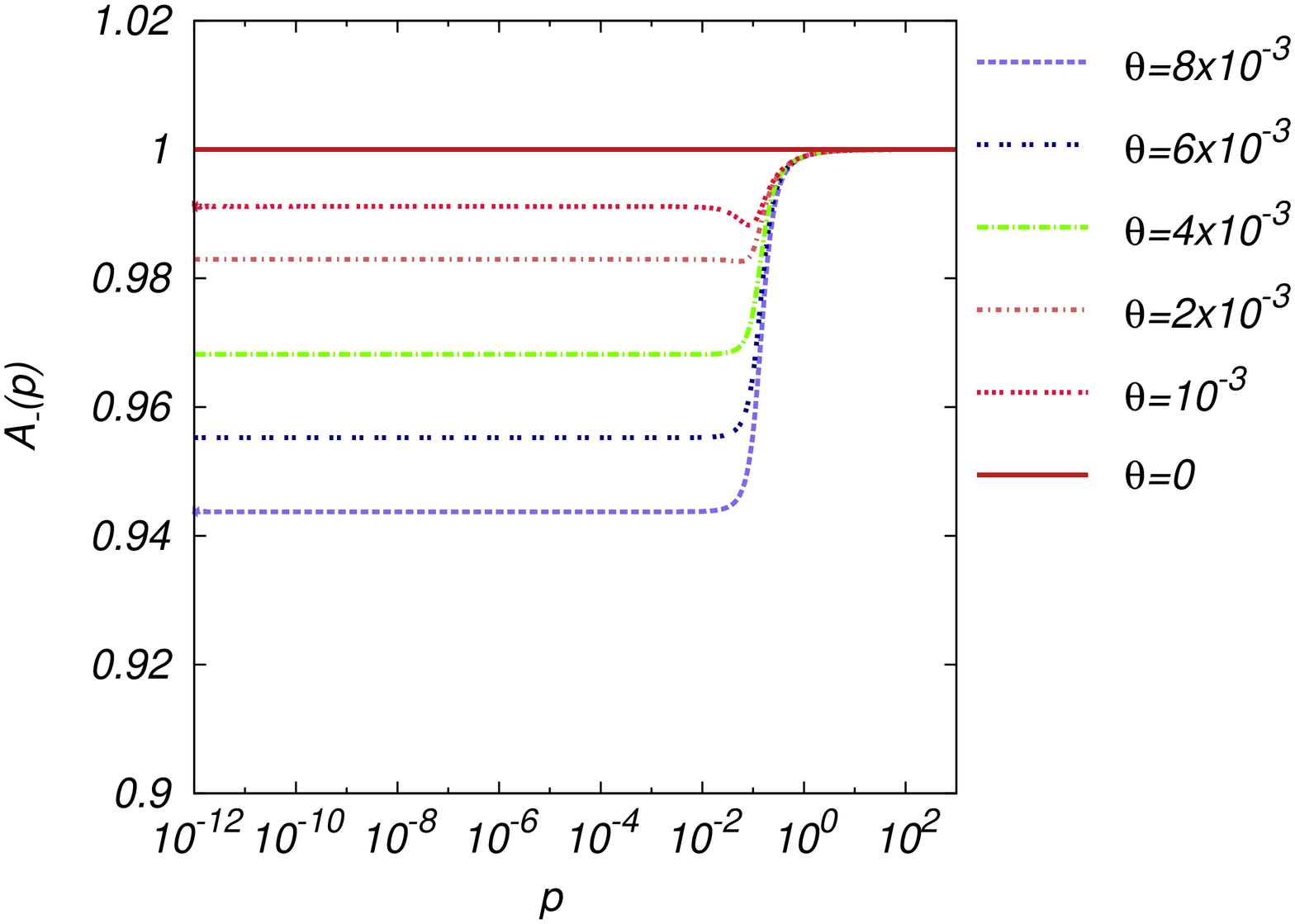}
\caption{Functions $A_\pm(p)$ for various values of $\theta$.}
\label{figApm}
\end{center}
\end{figure}

With the chiral decomposition of the fermion propagator, Eq.~(\ref{fpp}), both in the left- and right-handed
sector, the SDE corresponds to a matrix equation which can be converted into a system of scalar equations
for $A_\pm(p)$ and $B_\pm(p)$ after multiplying it by $1$ and $\slashed{p}$ and taking the trace. In the
limit $m_\pm=0$, this leads to the decoupled system of equations
\begin{eqnarray}
 A_{\pm}(p) &=& 1+
 \frac{\alpha}{ \pi^{2} p^{2}}\int d^{3} k \ \sigma^{V}_\pm(k) 
\frac{(k\cdot q)(p\cdot q)}{q^{2}(q^{2}+\theta^{2})} 
\mp \frac{\alpha \theta}{ \pi^{2} p^{2}}\int d^{3} k \ \sigma_\pm^S(k)
\frac{(p\cdot q)}{q^{2}(q^{2}+\theta^{2})} \;,\nonumber \\
 B_{\pm}(p) &=& 
\frac{\alpha}{ \pi^{2}}\int d^{3} k \ \sigma_\pm^S(k)
\frac{1}{(q^{2}+\theta^{2})} 
\mp \frac{\alpha \theta}{ \pi^{2}}\int d^{3} k \ \sigma_\pm^V(k) 
\frac{(k\cdot q)}{q^{2}(q^{2}+\theta^{2})} ,
\label{system}
\end{eqnarray}
with $\alpha=1/(4\pi)$ as usual.
Because of the effects of the CS term, the $A_\pm(p)$ functions become not trivial. Thus, in covariant
gauges these never decouple from the equations for $B_\pm(p)$ as long as $\theta\ne 0$. We numerically solve
the above systems of equations in each sector varying $\theta$.

\begin{figure}
\begin{center}
\includegraphics[width=0.35\textwidth, angle=270]{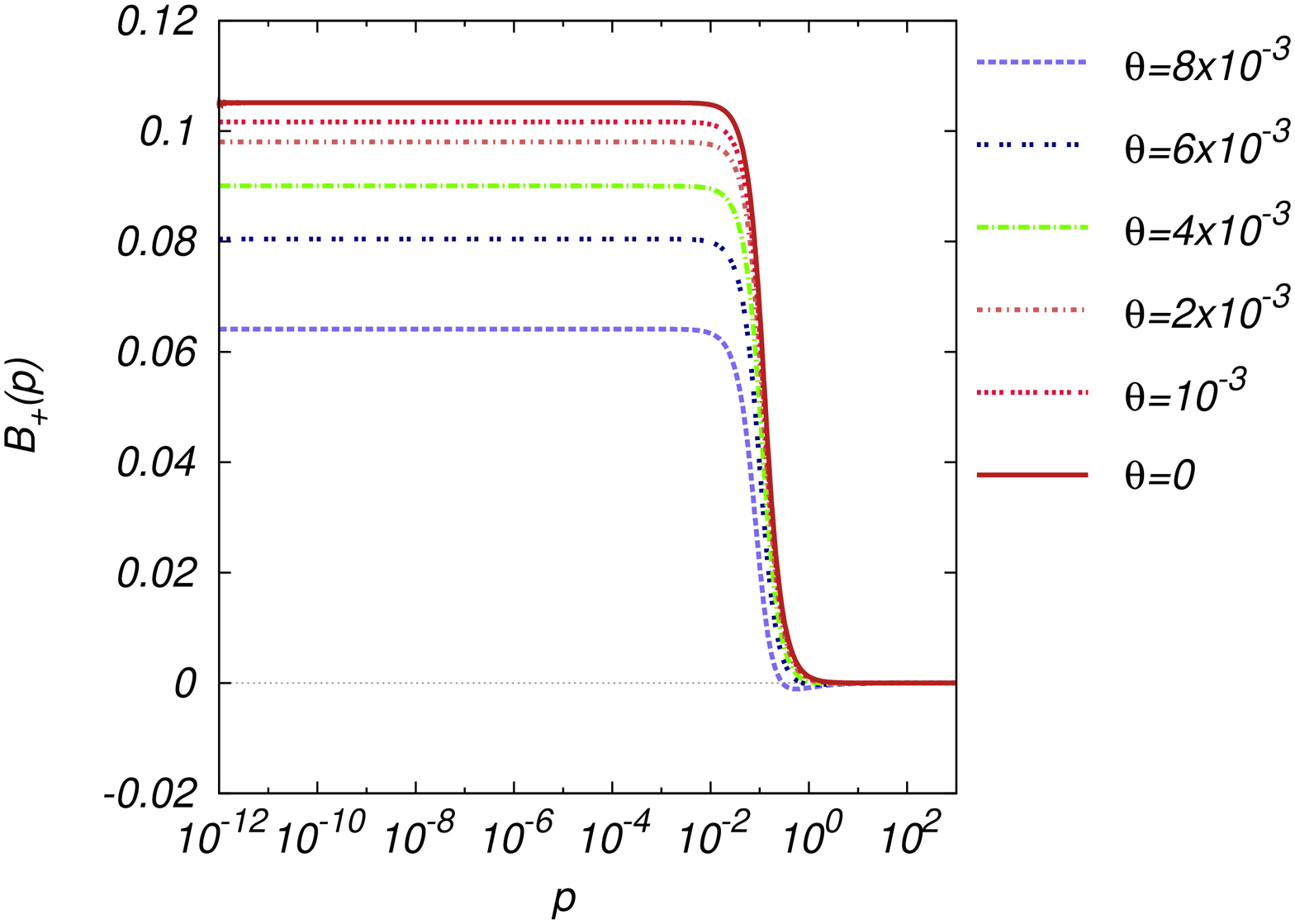}\includegraphics[width=0.35\textwidth,
angle=270]{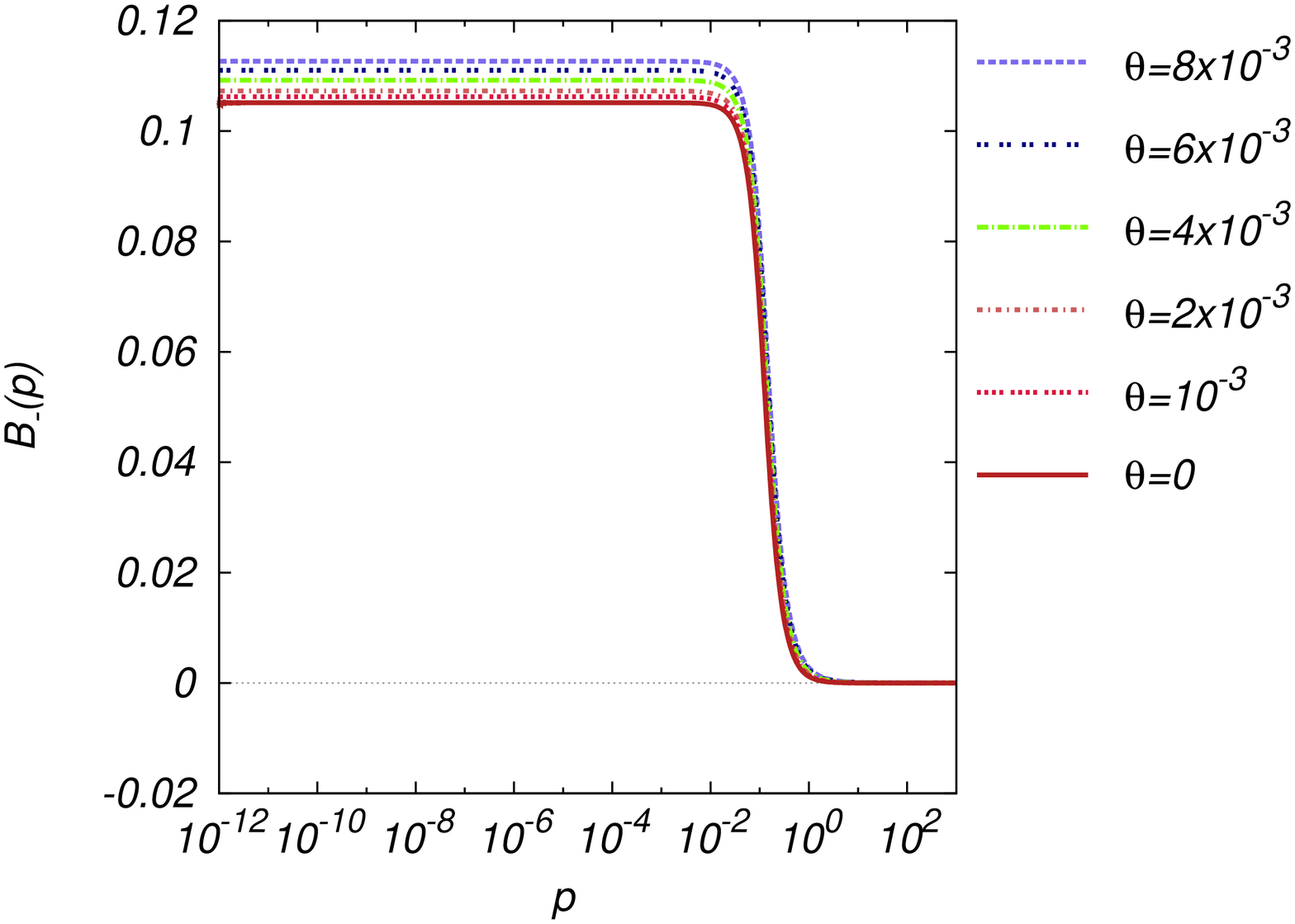}
\caption{Functions $B_\pm(p)$ for various values of $\theta$.}
\label{figBpm}
\end{center}
\end{figure}

\begin{figure}
\begin{center}
\includegraphics[width=0.35\textwidth, angle=270]{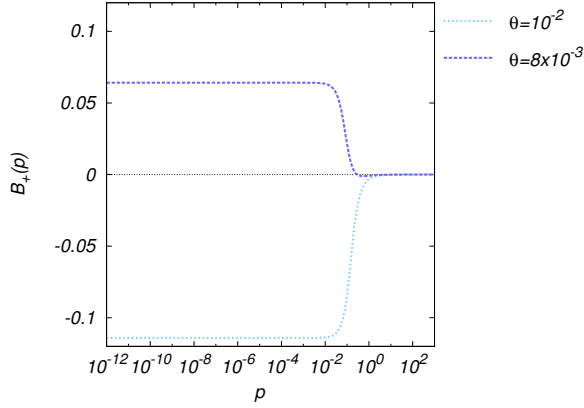}
\caption{Critical behavior of $B_{+}(p)$ for $\theta\simeq 8\times 10^{-3}$.}
\label{figBneg}
\end{center}
\end{figure}

In Fig.~\ref{figApm} we show the functions $A_\pm(p)$ for different values of $\theta$. The small deviation
from the tree-level value for small momentum in each case is due to the CS term. As $\theta$ increases, both
functions exhibit a plateau for small $p$ and smoothly tend to 1 in the ultraviolet. The deviation is more
pronounced in $A_-(p)$. Figure~\ref{figBpm} shows the result for $B_\pm(p)$. We observe that the height of
the plateau of $B_+(p)$ at small momentum decreases from the parity-even result as $\theta$ gets bigger, but
the height of the plateau of $B_-(p)$ exhibits the opposite behavior. The effect is more visible in
$B_+(p)$. In fact, there is a critical value $\theta_c\simeq8\times10^{-3}$ above which $B_+(p)$ strongly
changes its behavior and the plateau becomes negative, as shown in Fig.~\ref{figBneg}.

\begin{figure}
\begin{center}
\includegraphics[width=0.35\textwidth, angle=270]{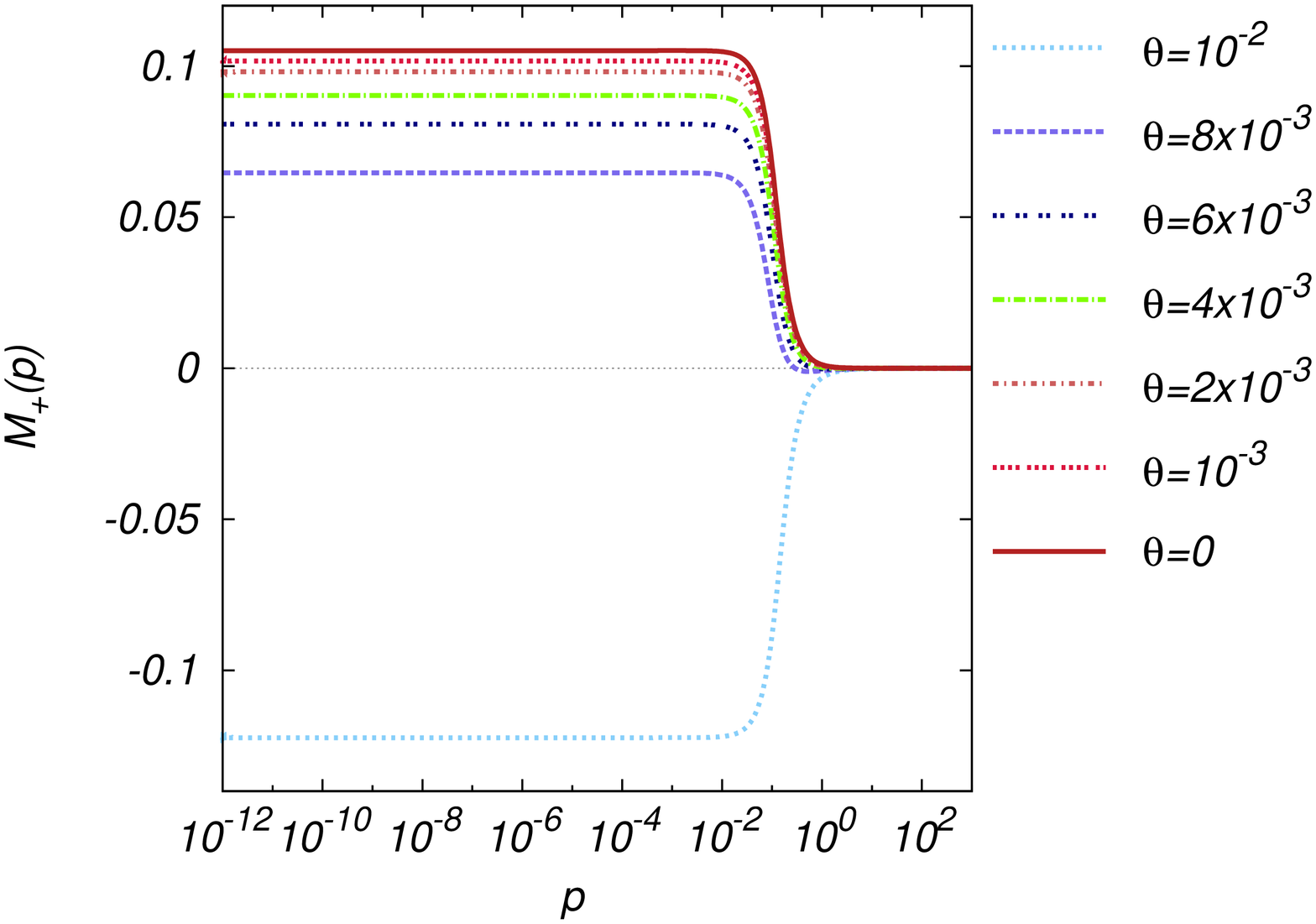}\includegraphics[width=0.35\textwidth,
angle=270]{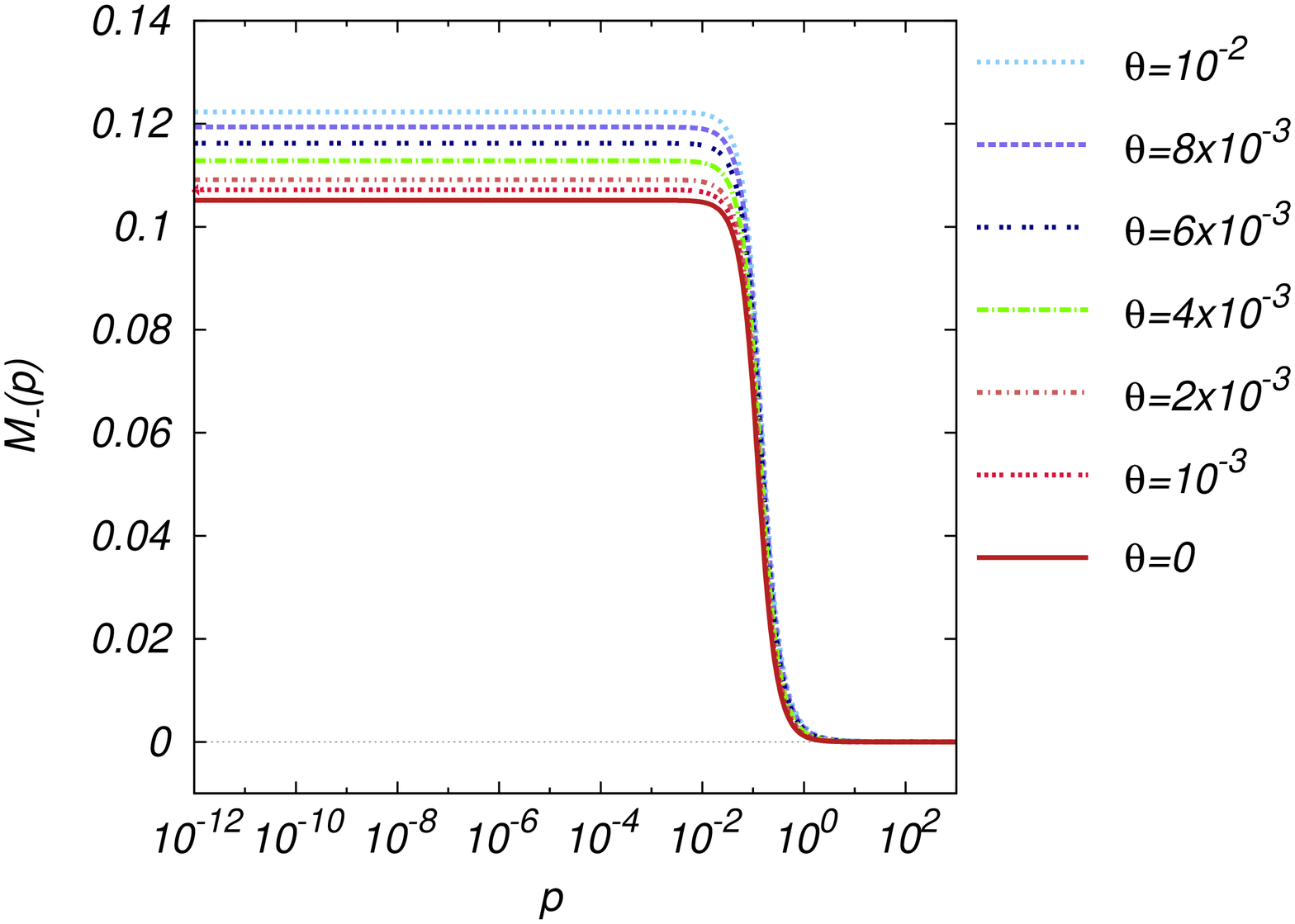}
\caption{Mass functions $M_\pm(p)$ for various values of $\theta$.}
\label{figMpm}
\end{center}
\end{figure}

\begin{figure}[h]
\begin{minipage}{14pc}
\includegraphics[width=12pc,angle=270]{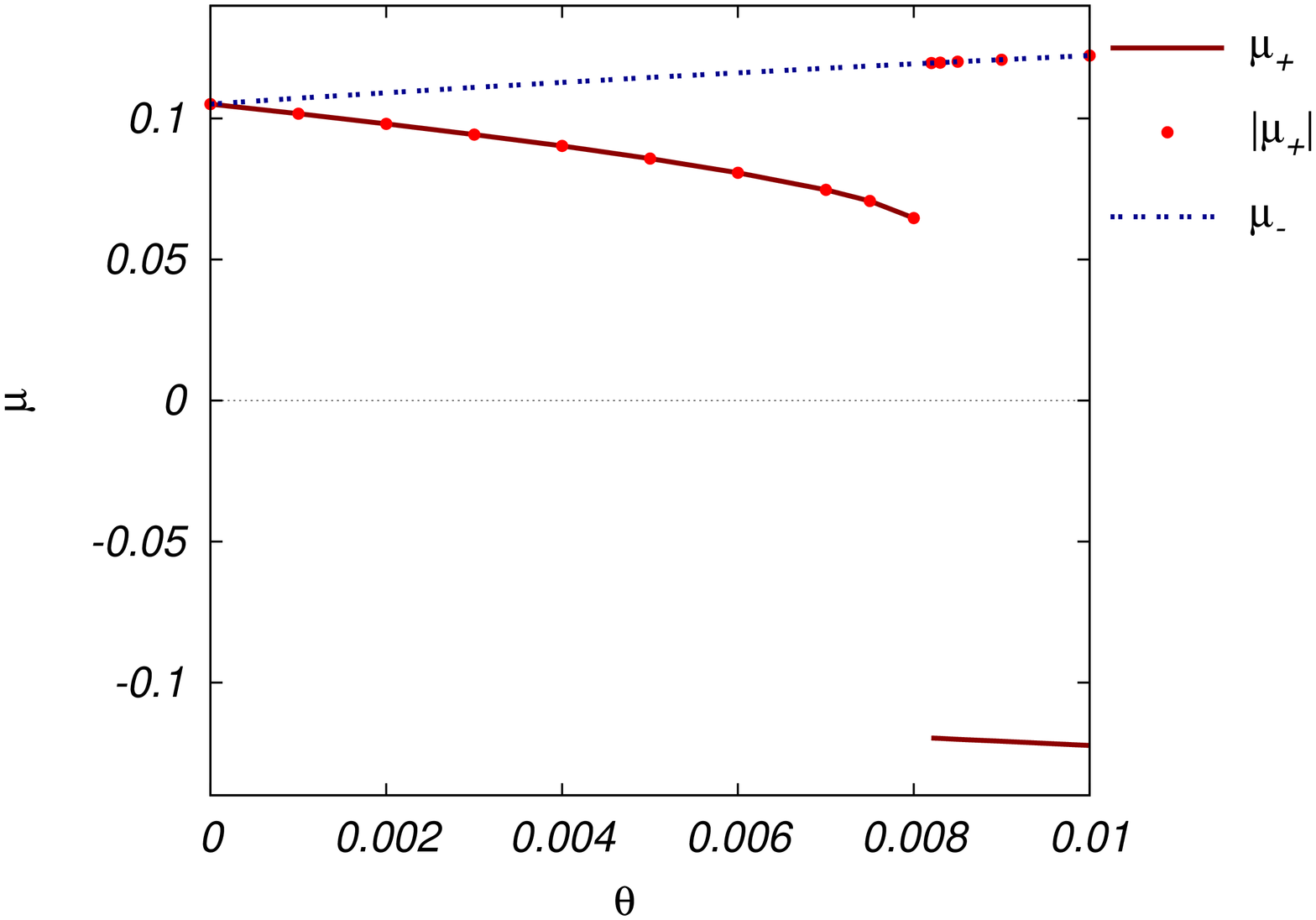}
\caption{$\mu_\pm$ as a function of $\theta$.}
\label{figmupm}
\end{minipage}\hspace{2pc}%
\begin{minipage}{14pc}
\includegraphics[width=12pc,angle=270]{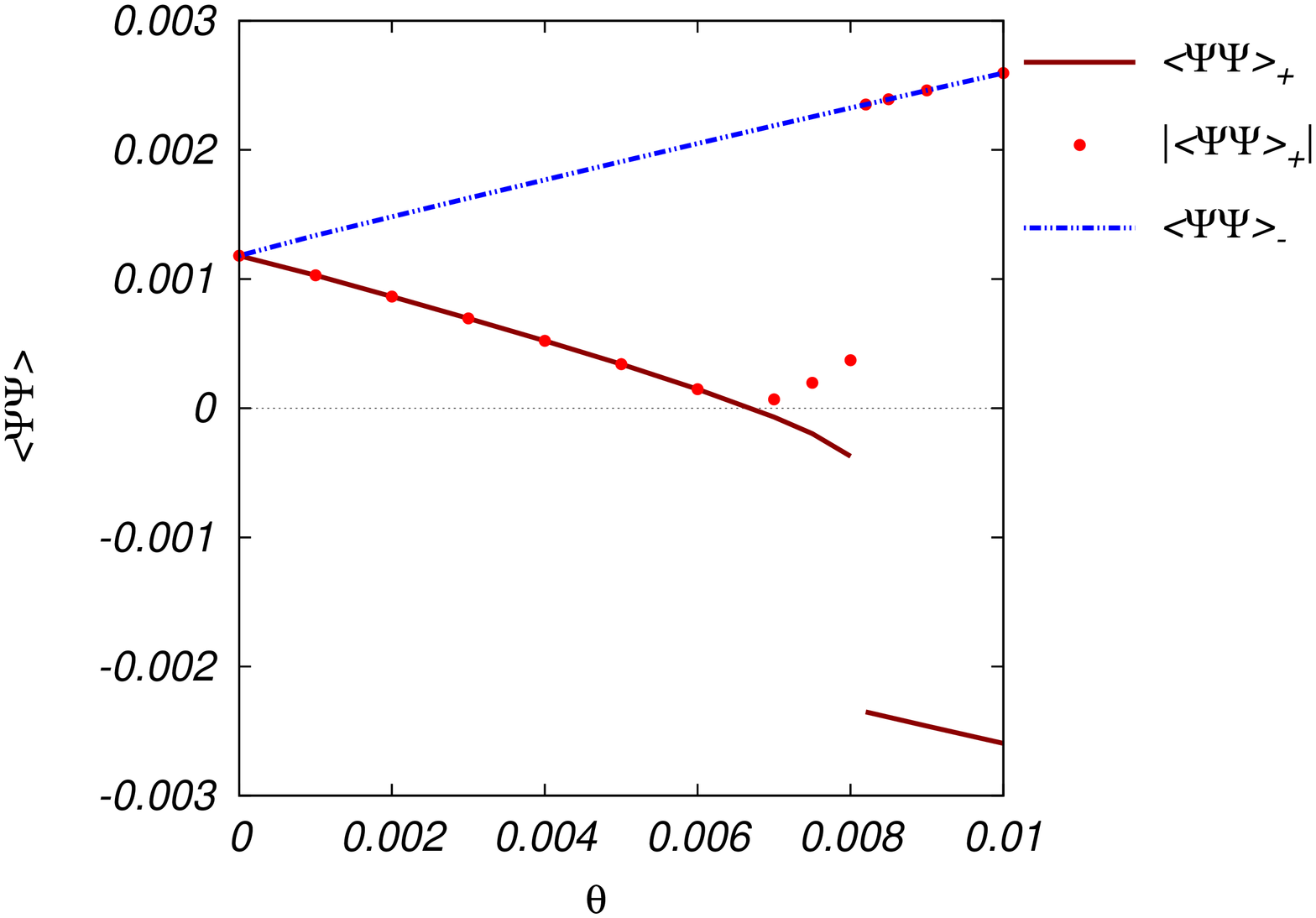}
\caption{ Chiral condensates as a function of $\theta$.}
\label{figCondpm}
\end{minipage} 
\end{figure}

This sudden drop has strong implications regarding chiral symmetry restoration in the
model~\cite{KondMarPRL,KondMarPRD}. In order to illustrate this, in Fig.~\ref{figMpm}, we show the mass
functions $M_\pm(p)=B_\pm(p)/A_\pm(p)$ for different values of $\theta$. At $\theta_c$, the plateau of
$M_+(p)$ also drops to negative values, whereas for $M_-(p)$, it continues to increase with $\theta$. To
ensure that these results are physically sensitive, we study the asymptotic behavior of $M_\pm(p)$. In the
infrared, the height of the plateau can be considered as an order parameter for dynamical chiral symmetry
breaking. In this connection, in Fig.~\ref{figmupm} we draw the dependence of $\mu_\pm=M_\pm(0)$ as a
function of $\theta$ below and above criticality. This parameter can be regarded as the dynamical mass of
the corresponding fermion species. We observe that the role of the CS coefficient is to remove the mass
degeneracy between fermion species as long as $\theta<\theta_c$. There is a light and a heavy species. At
$\theta_c$, however, there is a drastic change in this behavior, the light species develops a negative
mass, which in absolute value is the same as its heavy cousin, as can be appreciated in the dotted curve in
the same graph. This implies that above this $\theta_c$, the would-be parity preserving mass
$\mu_e=(\mu_++\mu_-)/2$ vanishes, and hence chiral symmetry is restored~\cite{KondMarPRL,KondMarPRD}.
Nevertheless, we want to emphasize that in this model, $\mu_e$ does not correspond to a pole in a propagator
and hence to a physical mass. Physical masses $\mu_\pm$ are generated for arbitrarily large values of
$\theta$.

The discontinuous behavior of the dynamically generated masses suggest a first order phase transition for
the dynamical breaking of chiral symmetry. Whether there is chiral symmetry restoration can better be seen
from the behavior of the condensates, which are the true order parameters. These can be extracted from the
solution to the SDEs as
\begin{eqnarray}
<\bar\psi \psi>_{\pm}=\frac{1}{\pi^2}\int_0^\infty dk k^2 \sigma_\pm^S(k)\;.
\end{eqnarray}
The results are shown in Fig.~\ref{figCondpm}. Notice that in the parity preserving case, $\theta=0$, the
value of the condensate for each species is half the value of the ordinary
condensate~\cite{Craigconf2,BRfbs}. The reason is that when fermion species are degenerate in mass, each one
of them, corresponding to a fermion pairing mode $\bar\psi_\downarrow\psi_\uparrow$ or
$\bar\psi_\uparrow\psi_\downarrow$ (arrows indicate spin orientation), contributes one half to the parity
preserving condensate.

\begin{figure}
\begin{center}
\includegraphics[width=0.35\textwidth, angle=270]{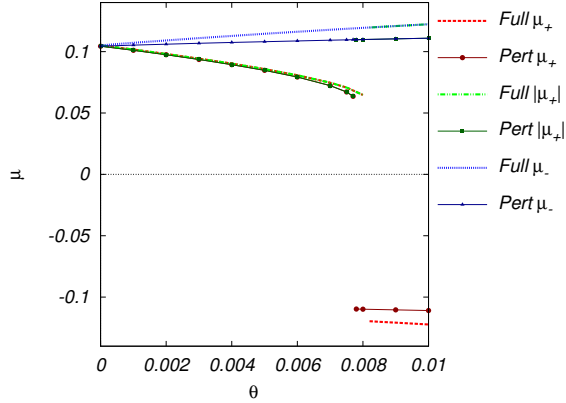}
\caption{Comparison of the dynamically generated fermion masses for the full system~(\ref{system}) and the
perturbatively reduced system (\ref{pertsystem}).}
\label{figpert}
\end{center}
\end{figure}

Finally, let us consider wavefunction renormalization effects. Regarding the wavefunction renormalization in
the dynamically generated mass functions $M_\pm(p)$, we compare the solutions of the full systems in the
previous subsection, Eq.~(\ref{system}), against the solution for the simplified equations resulting from
the perturbation theory motivated ansatz $A_\pm(p)=1$, namely
\begin{eqnarray}
 B_{\pm}(p) &=& 
\frac{\alpha}{ \pi^{2}}\int d^{3} k \frac{B_{\pm}(k)}{k^{2}+B_{\pm}^{2}(k)} \frac{1}{(q^{2}+\theta^{2})}
\mp \frac{\alpha \theta}{ \pi^{2}}\int d^{3} k \frac{1}{ k^{2}+B_{\pm}^{2}(k)} \frac{(k\cdot q)}{q^{2}(q^{2}+
\theta^{2})} \;.
\label{pertsystem}
\end{eqnarray}
The behavior of the solutions in this truncation is qualitatively similar to that of the complete
system~(\ref{system}). The critical behavior of the dynamical masses is not washed out by the effect of
wavefunction renormalization, as can be seen in Fig.~\ref{figpert}, where a comparison of $\mu_\pm$ and
$|\mu_+|$ between the full result and the perturbatively motivated one is shown. The behavior is virtually
the same in both truncation schemes. The critical value $\theta_c^{pert}$ is of the same order of $\theta_c$
and thus the effects of $A_\pm(p)\approx 1$ are not significant. This means that even with the inclusion of
the CS term, the rainbow-ladder truncation is still reliable in Landau gauge.

\section{Confinement}

\begin{figure}
\begin{center}
\includegraphics[width=0.35\textwidth, angle=270]{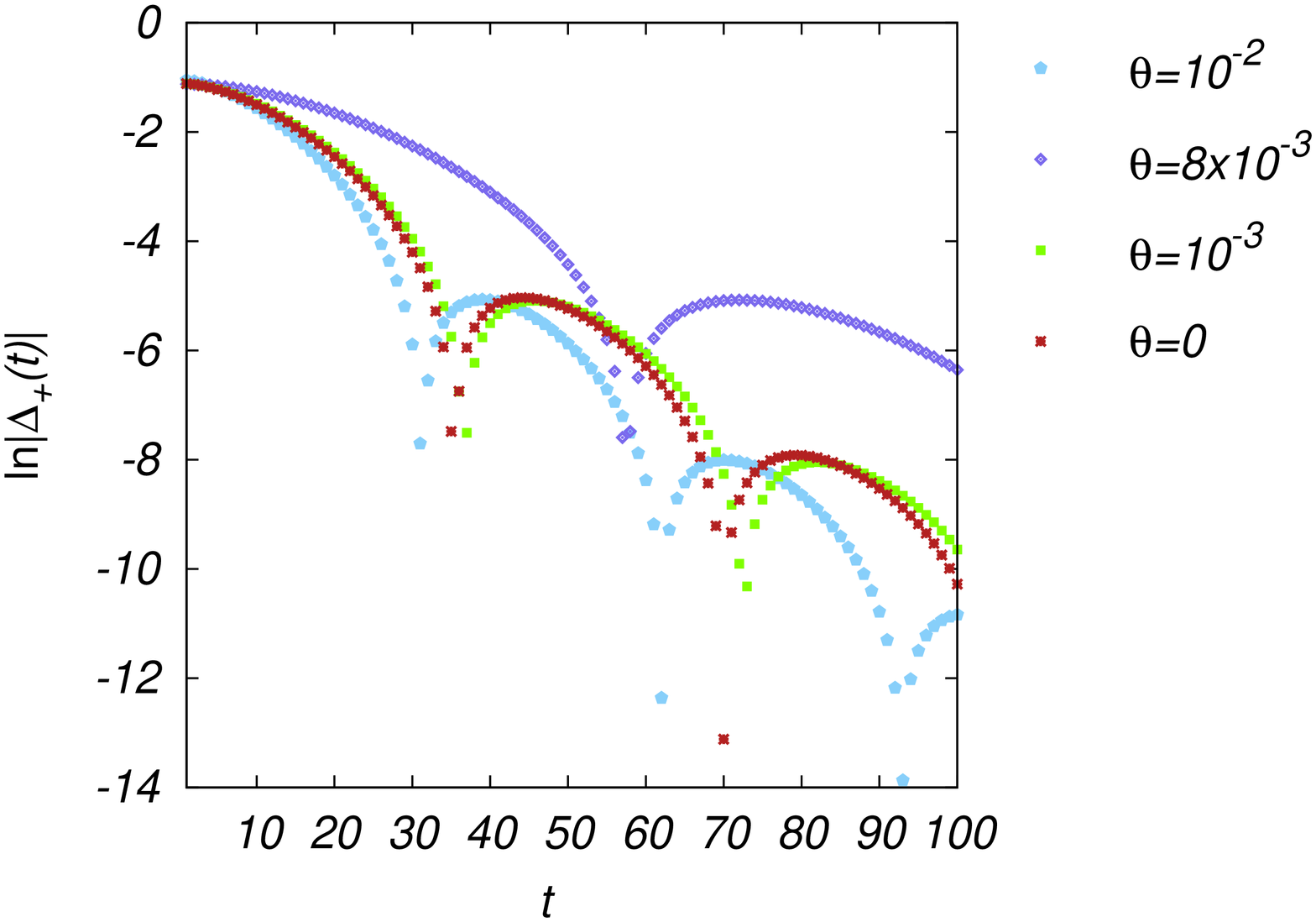}\includegraphics[width=0.35\textwidth,
angle=270]{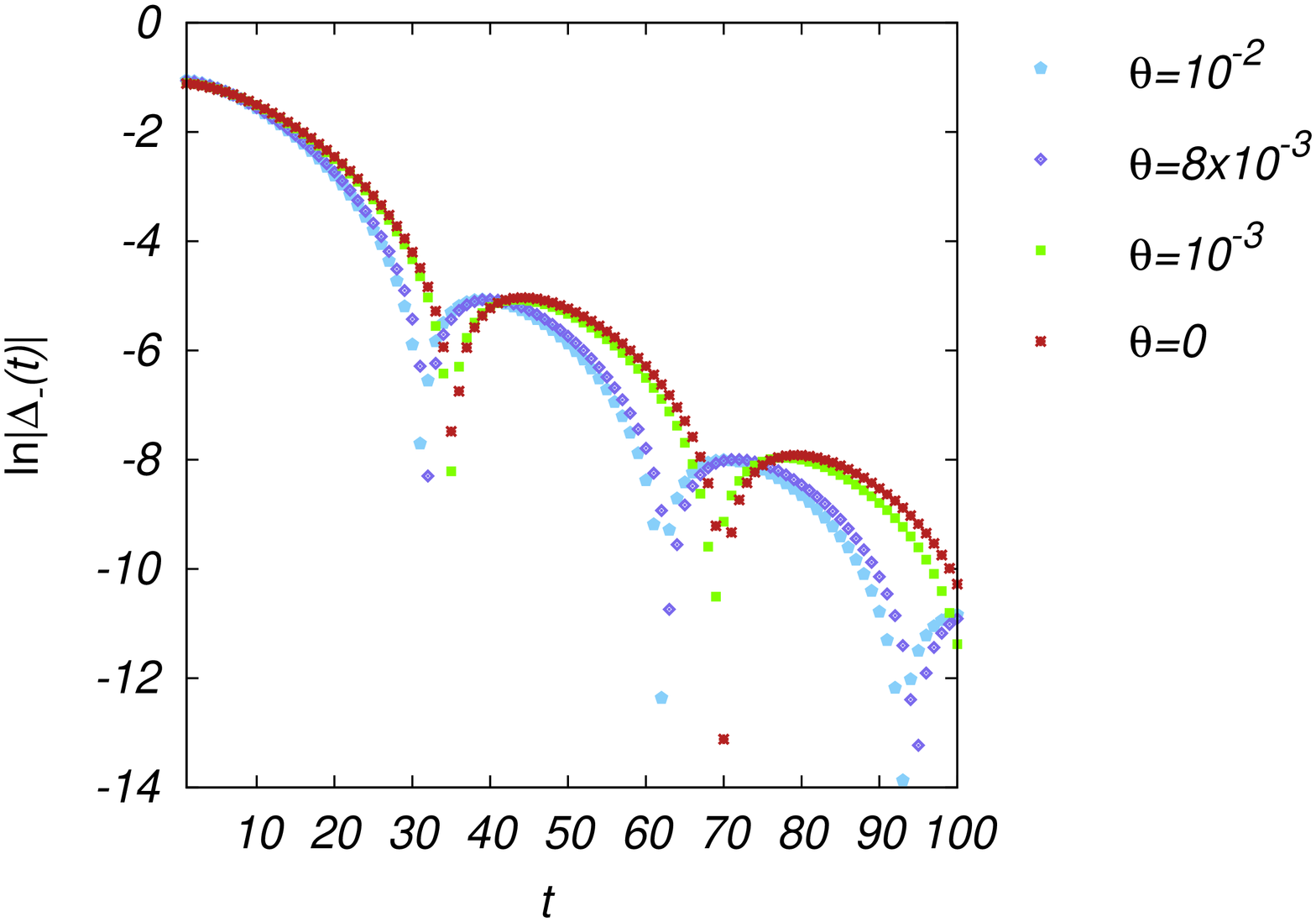}
\caption{Confinement tests for the spatially averaged Schwinger functions.}
\label{figConfOsc1}
\end{center}
\end{figure}

In ordinary quenched QED$_3$, there is confinement. Here we are interested in the extent up to which this
scenario persist in the presence of a CS term. Whether a solution to the SDE supports confinement can be
tested by means of the violation of the Osterwalder-Schrader axiom of reflection positivity, which states
that the spatially averaged Schwinger functions
\begin{equation}
\Delta_\pm(t)=\int d^2 x \int \frac{d^3 p}{(2\pi)^3}e^{i(t p_{o}+x\cdot p)}\sigma_\pm^S(p)\;,\label{spati}
\end{equation}
should be positive definite if they are related to a stable asymptotic state. In our case, 
we construct the functions $\Delta_\pm(t)$ inserting the solutions for the SDE in the above expression, upon
which we perform a confinement test. In Fig.~\ref{figConfOsc1} we plot the logarithm of the absolute value
of each of these functions. The oscillatory behavior inferred from the pronounced peaks in the graphs
reveals that even in the presence of a CS term, quenched QED$_3$ exhibits confinement. This behavior
corresponds to two complex conjugate mass like singularities with complex masses $m= a \pm ib$ for each
species, which fit to 
\begin{equation}
\Delta_\pm(t)\simeq C_\pm e^{-a_\pm t}cos(b_\pm t+\delta_\pm)\;.
\end{equation}
Thus, confinement is observed in our model.

\begin{figure}
\begin{center}
\includegraphics[width=0.35\textwidth, angle=270]{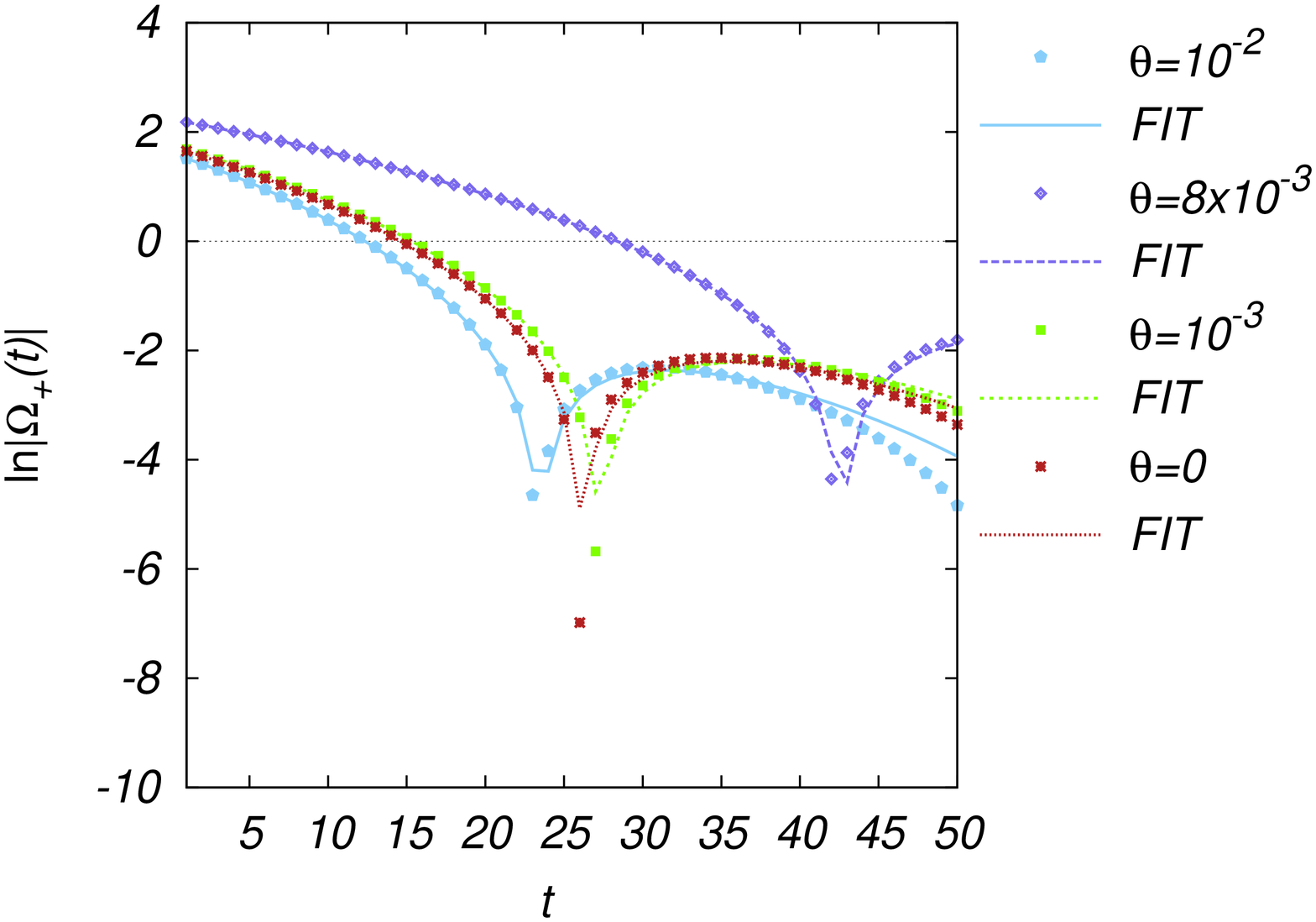}
\includegraphics[width=0.35\textwidth,angle=270]{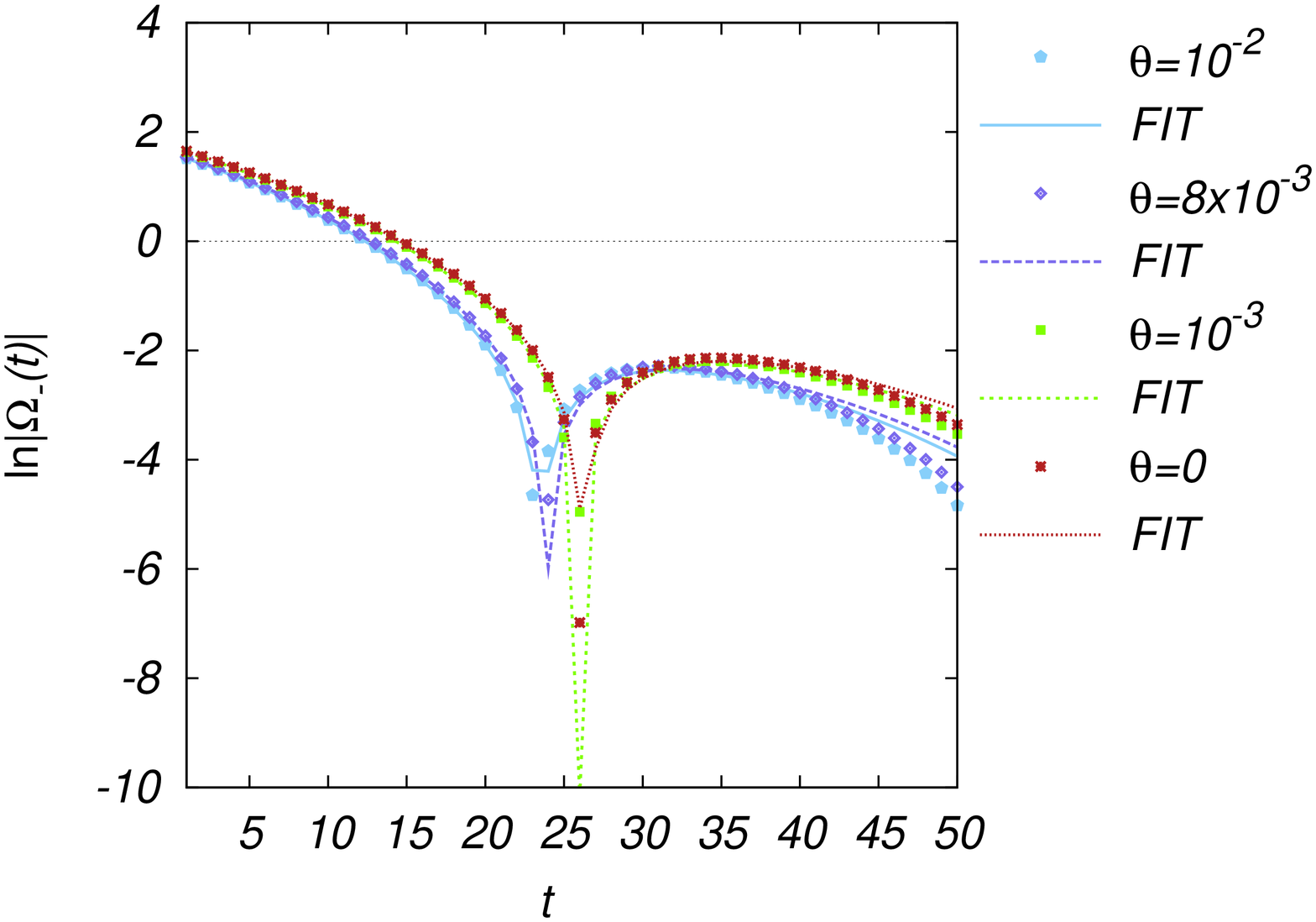}
\caption{Spatially averaged Schwinger functions involving the vector parts of the fermion propagator.}
\label{SIGMAV}
\end{center}
\end{figure}

Following Ref.\cite{raya-saul}, we perform a similar analysis on the vector parts of the propagator,
$\sigma_{\pm}^{V}$. We define new set of spatially averaged Schwinger functions,
\begin{equation}
\Omega_\pm(t)=\int d^2 x \int \frac{d^3 p}{(2\pi)^3}e^{i(t p_{o}+x\cdot p)}\sigma_{\pm}^{V}(p),\label{vpati}
\end{equation}
and insert again the solutions found in previous sections. 
In Fig.~\ref{SIGMAV} we draw the logarithm of the absolute values of $\Omega_\pm(t)$. Again, the
oscillations hint confinement. Moreover, we obsverve that $\Omega_\pm(t)$ can be fitted according to
\begin{equation}
 \Omega_{\pm}(t)\simeq C'_{\pm}e^{-{a'}_\pm t}cos({b'}_\pm t+{\delta'}_\pm)\;,
\end{equation}
which also unveils the complex pole mass structure of the propagator.

\section{Conclusions}

We have studied the dynamical generation of masses and confinement in Maxwell-Chern-Simons QED$_3$. There
exists two types of fermions within the four-component spinor formalism. These species are non-degenerated
in mass. The origin of these physical masses is two-fold: On the one hand there is a parity-preserving
contribution $m_e$ coming from dynamical chiral symmetry breaking, on the other hand, there is the
CS-induced Haldane mass term $m_o$. We have explored the consequences of confinement in our model. 

For $\theta>\theta_{c}$ the masses for each species are equal in magnitude, $\mu{+}=-\mu_{-}$, and we do
not generate the ordinary mass $\mu_{e}=(\mu_{+}+\mu{-})/2$ that preserves parity. This is the reason why
there is restoration of chiral symmetry and we only generated masses induced by the CS term. The phase
transition for the mass $ \mu_ {+} $ is of first order. Regarding confinement, according to tests performed
with $\sigma^{V}$ and $\sigma^{S}$, we observe that quenched QED$_3$ with a CS term also shows this feature,
regardless the value of the CS coefficient. This is in contrast to the case in which vacuum polarization
effects are considered in the leading order of the $1/N$ approximation~\cite{raya-saul}, where there is
deconfinement for arbitrary finite values of $\theta$.

\ack
We acknowledge support from SNI, CIC, and CONACyT grants through projects 4.22, 82230, and 50744-F,
respectively.

\end{document}